\documentclass[aps,preprint,groupedaddress]{revtex4}
\usepackage{graphicx}
\usepackage{dcolumn}
\usepackage{bm}

\begin{document}
\title{Electric readout of magnetization dynamics in a ferromagnet-semiconductor system}
\author{\L.~Cywi{\'n}ski}\email{cywinski@physics.ucsd.edu}
\author{H.~Dery}
\author{L.~J.~Sham}
\affiliation{Department of Physics, University of California San
Diego, La Jolla, California, 92093-0319}
\date{\today}
%%%%%%%%%%%%%%%%%%
\begin{abstract}
We apply an analysis of time-dependent spin-polarized current in a semiconductor channel at room temperature to establish how the magnetization configuration and dynamics of three ferromagnetic terminals, two of them biased and third connected to a capacitor, affect the currents and voltages. In a steady state, the voltage on the capacitor is related to spin accumulation in the channel. When the magnetization of one of the terminals is rotated, a transient current is triggered. This effect can be used for electrical detection of magnetization reversal dynamics of an electrode or for dynamical readout of the alignment of two magnetic contacts. 
\end{abstract}
\maketitle
%%%%%%%%%%%%%%%%%%
Integration of non-volatile magnetic memory into semiconductor electronics is one of the goals of spintronics. Current magnetic random-access memory (MRAM) implementations are based on all-metallic systems,\cite{Tehrani_IEEE00} maintaining a physical separation between the memory and the logic parts in computer architectures. Several theoretical proposals of spin transistors involving semiconductors have been put forth, \cite{DattaDas_APL90,Schliemann_PRL03,Ciuti_APL02,Fabian_PRB04,Flatte_APL03} but the experimental verification is yet to be made. Recently, we have proposed a system consisting of three ferromagnetic metal contacts on top of a planar semiconductor channel, \cite{Dery_MCT_PRB06} in which the flexibility provided by a third terminal allows to directly express the magnitude of spin accumulation in the semiconductor channel by electrical means.
This system relies on currently available parameters of metal/semiconductor spin injecting structures involving Fe and GaAs . \cite{Hanbicki_APL02,Hanbicki_APL03,Adelmann_PRB05,Crooker_Science05}

In this Letter, we explore the possibility of a {\it dynamical} read-out scheme in a three-terminal system in Fig.~1 based on a time-dependent analysis of the {\it lateral} spin diffusion under the ferromagnetic contact. 
The lateral scale of the planar structure is set by the spin diffusion length $L_{sc}$, about 1~$\mu$m in GaAs at room temperature, 
the distance within which the electron spin polarization is preserved. 
The ferromagnetic terminals have collinear magnetizations.
Bias is applied between the left (L) and the middle (M) contacts. The right contact (R) is connected to a capacitor $C$ which blocks the current in steady state. The voltage on the capacitor depends on the alignment of L and M magnetizations as well as on the spin selective properties of the R terminal.  This is known as the non-local spin-valve effect \cite{Johnson_Science93,Jedema_Nature01,Ji_APL06}. 
In the following, we fix the M terminal magnetization, which  can be realized by using an antiferromegnetic capping layer.~\cite{Nogues_JMMM99} 
To change the magnetizations of L and R terminals the planar structure is augmented by a set of current-carrying lines known from MRAM devices.~\cite{Tehrani_IEEE00}
We discuss two possible modes of operation for this system. 
In the first mode, the magnetization of L contact is perturbed, and its dynamics is driving a current in the R contact. This leads to the possibility of an all-electrical measurement of magnetization reversal. In the second mode, the L magnetization represents a bit of memory, and the rotation of the R contact triggers a transient current, the magnitude of which is related to the relative alignment of L and M magnetizations. 

\begin{figure}
\includegraphics[height=8cm]{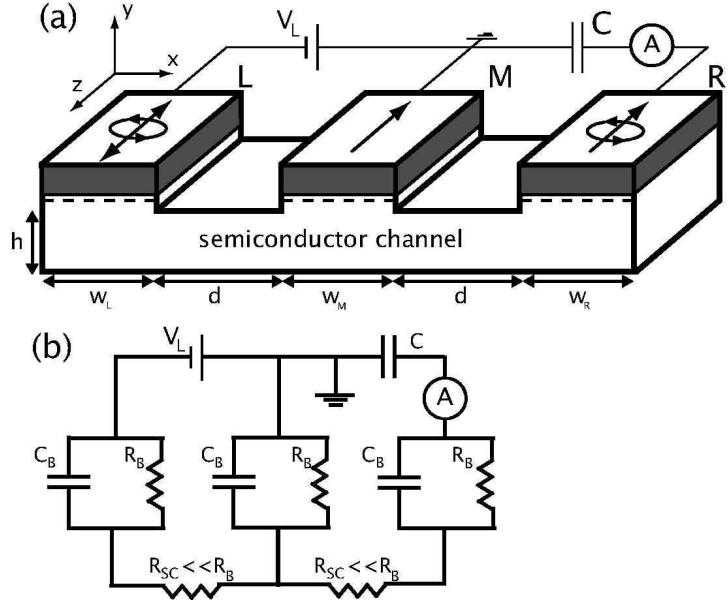}
\caption{(a) Proposed system: the structure is grown on top of an insulating layer as a mesa. The magnetization directions are manipulated by wire strips (not shown) above and below the structure, like in MRAM cell. Current in the R contact is measured. In the calculations we use h$=$$100$ nm, w$_{L}$$=$w$_{M}$$=$w$_{R}$$=$$400$ nm, d$=$$200$ nm and length in the $z$ direction of 2 $\mu$m. 
(b) The equivalent circuit diagram (spin-independent): $C_{B}$ is barrier capacitance, $R_{B}$ and $R_{SC}$ are barrier and semiconductor channel resistances, respectively . See the text for description of spin effects. 
}
\end{figure}

The spin accumulation in the channel and its connection to the alignment of the contacts is crucial for understanding of the system's operation. 
The current passing between L and M electrodes is spin polarized due to the spin selectivity of thin Schottky barriers.~\cite{Hanbicki_APL02,Hanbicki_APL03} The amount of spin accumulation (proportional to the spin splitting of electrochemical potential $\Delta \xi$ defined below) in the semiconductor depends on whether the L and M magnetizations are parallel (P) or anti-parallel (AP). In the P case, the excess injected spin population is easily drained from the channel, while in the AP case opposite spins are more efficiently injected and extracted, leading to much larger spin accumulation. Using the notations in Fig.~1
the effective length of the active channel covered by the L and M
terminals is $l$$\approx$$d$$+$$w_L$$+$$w_M$.   
Beneath the R contact, we then have from Ref.~\onlinecite{Dery_MCT_PRB06}  approximately $\Delta \xi_{AP}/\Delta \xi_{P}$$\approx$$(2L_{sc}/l)^2$. 
In the steady state, when no current is flowing through R electrode, its electrochemical potential $\mu_{R}$ (having spin splitting negligible compared to the splittings in the semiconductor) depends on the spin accumulation beneath the contact, and on the direction of R magnetization relative to the reference direction of M magnet. The boundary condition \cite{Yu_Flatte_long_PRB02,Dery_lateral_PRB06} connecting the electrochemical potential $\xi_{s}$ (spin $s$$=$$\pm$)  with the spin current  $j_{s}$ at the interface is $ej_{s}$$=$$G_{s}(\mu_{R}-\xi_{s})$, where $G_{s}$ is the barrier conductance for spin $s$.
Using this, the requirement of zero net current gives
\begin{equation}
\mu_{R} = \xi +  \frac{G^{R}_{+}-G^{R}_{-}}{G^{R}_{+}+G^{R}_{-}} \, \frac{\Delta \xi}{2} \,\, ,  	 \label{eq:steady}
\end{equation}
where $\xi$ is the mean potential beneath R contact and $\Delta \xi$ is its spin splitting. When the L magnetization is rotated, $\Delta \xi$ changes, and when R magnet is switched, $G^{R}_{+}$ and $G^{R}_{-}$ trade places. In both cases, perturbation of one of the magnets leads to transient currents charging the capacitor.
The values of steady-state $\mu_R$ for different contact alignments are illustrated in Fig.~2 (dashed lines) with respect to the spin-resolved potentials in the channel beneath the R contact (solid lines). The perturbation of either R or L magnet changes $\mu_{R}$ between these values. 
If the RC time constant of the entire circuit is shorter than a time-scale of magnetization dynamics, it is possible to trace out the magnetization dynamics by electrical means. If $\Delta \xi$ is unchanged (when only R is rotated), then the signal is expected to differ strongly in magnitude depending of the L/M alignment, due to the ratio of spin splittings mentioned above. This leads to a dynamical readout of the L magnetization direction by rotating the R magnet.

\begin{figure}
\includegraphics[height=6cm]{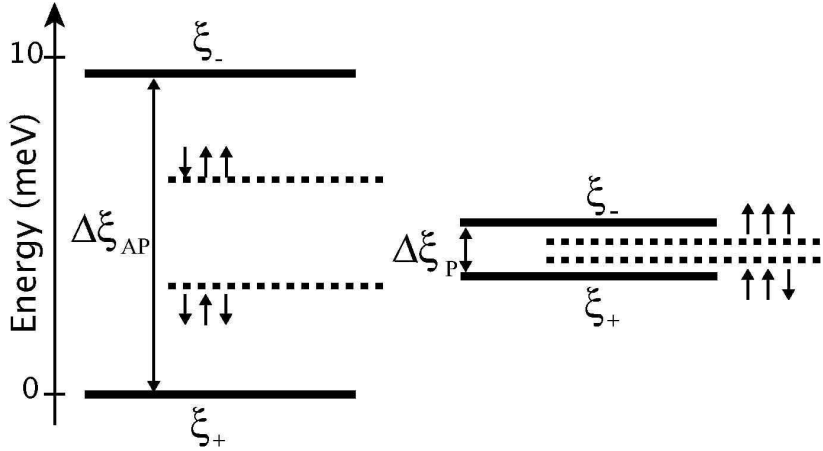}
\caption{ Spin accumulation under the R contact and voltage inside it for antiparallel (AP) or parallel (P)  magnetization alignment of the L and M terminals. Solid lines are the spin-dependent electrochemical potentials in the semiconductor channel beneath the R contact. Dashed lines are the  values of electrochemical potential $\mu_{R}$ in the R contact in the steady state, depending on the R direction, with arrows denoting the alignment of three magnets. }
\end{figure}

In contrast to previous treatments of time-dependent spin diffusion \cite{Rashba_APL02,Zhang_PRB02}, we treat the transport in a planar semiconductor by a method \cite{Dery_lateral_PRB06}, in which the effect of traversing under finite width of the metal contacts and barrier capacitance are taken into account. The electrochemical potential is defined \cite{Yu_Flatte_long_PRB02} for spin $s$$=$$\pm$ in a non-degenerate semiconductor as $\mu_{s}$$=$$2k_{B}Tn_{s}/n_{0}$$-$$e\phi$, where $n_{s}$ is the non-equilibrium part of the spin density, $n_{0}$ is the free electron concentration, and $\phi$ is the electrostatic potential. 
The spin selectivity of the barrier is described by the finesse $F$$=$$(G_{+}-G_{-})/G$ with $G$$=$$G_{+}+G_{-}$. 
We introduce two dimensionless parameters to quantify, respectively, the total conductance of the barrier and its spin selectivity: 
$\alpha$$=$$2L_{sc}^{2}G/(\sigma h)$ and $\beta$$=$$\alpha F$, where $h$ is the thickness of the conducting channel (see Fig.~1) and $\sigma$ is the conductivity of the semiconductor. The $y$-average of $\mu_{s}$ over the thickness of the channel, denoted by $\xi_{s}$ is used in the transport equations.~\cite{Dery_lateral_PRB06} In terms of the splitting $\Delta \xi $$=$$ \xi_{+}-\xi_{-}$ and the mean $\xi $$=$$ (\xi_{+}+\xi_{-})/2$ we have the spin diffusion equation
\begin{equation}
\frac{\partial \Delta \xi}{\partial t} = D \frac{ \partial^{2}
\Delta \xi}{\partial x^{2}} + \frac{\beta_{i}(t)}{\tau_{s}}(\mu_{i}-\xi) -\frac{\alpha_{i}}{2\tau_{s}} \Delta \xi -
\frac{\Delta \xi}{\tau_{s}}, \label{eq:Dxi}
\end{equation}
where $\mu_i$ is the electrochemical potential in the $i^{th}$ ferromagnet and $\tau_{s}$ is the semiconductor spin relaxation time.  
To complete the equations for $\xi$ and $\Delta \xi$, we use the excellent approximation of quasi-neutrality in the channel at all times ($n_{+}+n_{-}$$=$$0$).
In steady state, the quasi-neutrality condition follows from the smallness of the ratio of Fermi screening length to spin diffusion length.~\cite{Hershfield_PRB97} In the time-dependent case, deviations form neutrality are screened out on the scale of the dielectric relaxation time $\tau_{d}=\epsilon\epsilon_{0}/\sigma$, which is $\sim$100 fs for the semiconductor channel in our case \cite{Smith_Semiconductors}. For the dynamics on longer time-scales (at least tens of picoseconds), we can assume that at every time-step the quasi-neutrality is preserved. Consequently, $\xi$ is proportional to $\phi$ and it satisfies the Laplace equation with von Neumann boundary conditions related to currents at the boundaries of the channel, which in the time-dependent case include also a displacement current connected with charging of the barrier capacitance $C_{B}$. The equation for $\xi$ in the channel is then: 
\begin{equation}
\frac{ \partial^{2} \xi}{\partial x^{2}} =
-\frac{\alpha_{i}}{2L_{sc}^{2}}(\mu_{i}-\xi) +\frac{\beta_{i}(t)}{4L_{sc}^2} \Delta \xi - \frac{c_{B}}{\sigma h}\frac{\partial}{\partial t} (\mu_{i}-\xi)  \,. \label{eq:xi}
\end{equation}
where $c_{B}$ is the barrier capacitance per unit area, and the right hand side of Eq.~(\ref{eq:xi}) is non-zero only under the contacts.

The barrier conductances $G_{s}$ refer to the two spin directions $s$$=$$\pm$ along the quantization axis parallel to the M magnetization. During the magnetization dynamics, we employ the barrier finesse $F(t)$ value proportional to the projection of the magnetization on the quantization axis, while keeping total $G$ constant. Thus, we neglect the effects of  ``mixing conductance'', \cite{Brataas} which are expected to be small for tunneling barriers. The magnetization dynamics of $i^{th}$ contact  translates into time-dependence of $\beta_{i}$, driving the spin diffusion in Eq.~(\ref{eq:Dxi}) and electric potential in the channel in Eq.~(\ref{eq:xi}). From $\xi_{s}$ we calculate the current $I_{R}$ flowing into the right contact and charging the capacitor $C$, and consequently the electrochemical potential of the R terminal $\mu_{R}$$=$$-eV_{R}$ changes according to  $dV_{R}/dt$$=$$I_{R}/C$.  

For the electrical tracing of L magnetization dynamics, both M and R magnets should be pinned in the same direction. In the case of the dynamical readout of L/M alignment, we need to write separately the memory bit (direction of L magnet) and read by rotating the R magnet. A proper choice of different coercivities of two magnets and magnetic field pulses should allow for separate addressing. The half-selection (unintentional perturbation of magnetization)  of L when rotating R should be diminished, in order not to mix the signal from the L dynamics with the readout of L alignment.

\begin{figure}
\includegraphics[height=8cm]{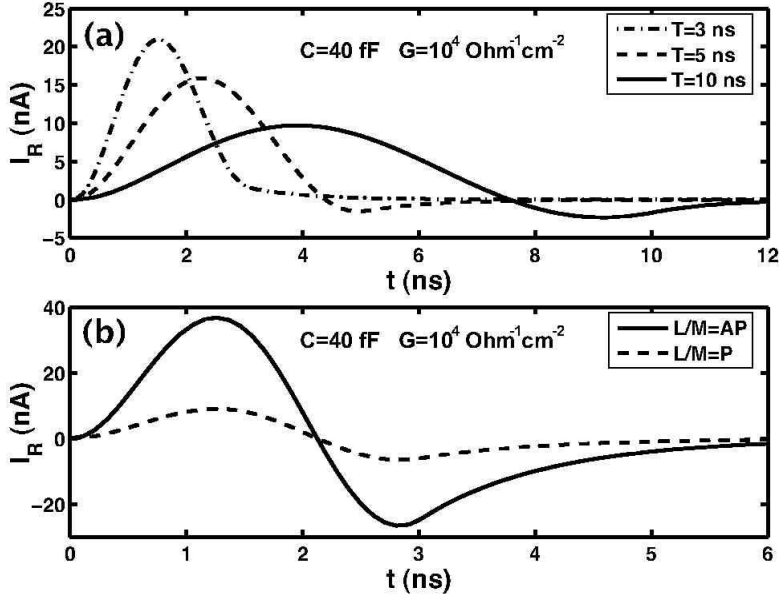}
\caption{ (a) R current signal for reversal of L magnetization occurring on time-scale of 3, 5 and 10 ns starting from AP alignment of L relative to M magnet. (b) R current signal for $2\pi$ rotation of R magnet for P and AP alignments of L and M magnets. The period of rotation is 3 ns. }
\end{figure}

For the calculations, we use the parameters of GaAs at room temperature: $\tau_{s}$$=$$80$ ps, doping $n$$=$$10^{16}$cm$^{-3}$, mobility $\nu$$=$5000 cm$^{2}$/Vs with corresponding diffusion constant $D$$=$$\nu kT/e$. The dimensions of the system are given in Fig.~1.  Beneath the barriers we assume a heavily doped profile \cite{Hanbicki_APL02} so that the Schottky barriers are thin ($<$$10$ nm), enabling spin injection \cite{Rashba_PRB00}. We employ the experimentally verified \cite{Hanbicki_APL03} spin selectivity $G_{+}/G_{-}$$=$$2$ and take the barrier conductance to be $G$$=$$10^{4}\,$$\Omega^{-1}$cm$^{-2}$. 
For such barriers of 1 $\mu$m$^{2}$ area and 10 nm thickness, $R_{B}$$=$$10$ k$\Omega$ and $C_{B}$$=$$10$ fF . 
The external capacitance is taken as $C$$=$$40$ fF, and the resulting RC time is about 1 ns. 
The applied voltage $V_L$ is $0.1$ V, and the ratio of forward to reverse biased $G$ is set to 2. 

In Fig.~3a we present the calculated $I_{R}$ induced by  reversal of the L magnet from AP to P alignment relative to M. 
In Fig.~3b the transient $I_{R}$ for the rotation of R occurring in 3 ns is shown.
While the average current is zero, the average power of the current pulse is much higher for the L/M$=$AP than for P. Two signals of such clearly different magnitude can be easily distinguished, provided that the stronger signal is above the noise level (dominated by Johnson noise in our system). In Fig.~3b the power of AP pulse is slightly above the noise power in 0.3 GHz bandwidth.

In summary, we have proposed a metal-semiconductor system in which the dynamics of one of magnets can be sensed electrically. This opens up a possibility for electrical detection of magnetization switching dynamics in buried structures, inaccessible to magneto-optical techniques. We have also discussed a possibility for dynamical read-out of magnetization direction of one of the terminals, which can be used for magnetic memory purposes.  
Our ideas are supported by calculations of time-dependent spin diffusion, taking into account realistic geometry of the structure. 
Further developments of including a scheme of  all-magnetic logic gate will be presented in upcoming publications.

We thank Parin Dalal for useful discussions. This work is supported by NSF DMR-0325599.


\begin{thebibliography}{99}
% 1
\bibitem{Tehrani_IEEE00} S. Tehrani, B. Engel, J. M. Slaughter, E. Chen, M. DeHerrera, M. Durlam, P. Naji, R. Whig, J. Janesky, J. Calder, IEEE Trans. Magn. {\bf 36}, 2752 (2000).
% 2
\bibitem{DattaDas_APL90}  S. Datta and B. Das, Appl. Phys. Lett \textbf{56}, 665 (1990).
% 3
\bibitem{Schliemann_PRL03} J. Schliemann, J. C. Egues and D. Loss, Phys. Rev. Lett. {\bf 90}, 146801 (2003).
% 4
\bibitem{Ciuti_APL02} C. Ciuti, J. P. McGuire, and L. J. Sham, Appl. Phys. Lett \textbf{81}, 4781 (2002).
% 5
\bibitem{Fabian_PRB04}  J. Fabian and I. {\u Z}uti{\'c}, Phys. Rev. B \textbf{69}, 115314 (2004).
% 6
\bibitem{Flatte_APL03} M. E. Flatt{\'e}, Z. G. Yu, E. Johnston-Halperin
and D. D. Awschalom, Appl. Phys. Lett \textbf{82}, 4740 (2003).
% 7
\bibitem{Dery_MCT_PRB06} H. Dery, \L.~Cywi{\'n}ski, and L. J. Sham, Phys. Rev. B \textbf{73} 161307(R) (2006).
% 8
\bibitem{Hanbicki_APL02} A. T. Hanbicki, B. T. Jonker, G. Itskos, G.
Kioseoglou, and A. Petrou, Appl. Phys. Lett. \textbf{80}, 1240
(2002).
% 9
\bibitem{Hanbicki_APL03} A. T. Hanbicki, O. M. J. van' t Erve, R. Magno, G.
Kioseoglou, C. H. Li, B. T. Jonker, G. Itskos, R. Mallory, M. Yasar,
and A. Petrou, Appl. Phys. Lett. \textbf{82}, 4092 (2003).
% 10
\bibitem{Adelmann_PRB05} C. Adelmann, X. Lou, J. Strand, C. J. Palmstrom, and P. A. Crowell, Phys. Rev. B \textbf{71}, 121301(R) (2005).
% 11
\bibitem{Crooker_Science05}  S. A. Crooker, M. Furis, X. Lou, C. Adelmann, D. L. Smith, C. J. Palmstrom, and P. A. Crowell, Science \textbf{309}, 2191 (2005).
% 12
\bibitem{Johnson_Science93} M. Johnson, Science \textbf{260}, 320
(1993).
% 13
\bibitem{Jedema_Nature01} F. J. Jedema, A. T. Filip,
and B. J. van Wees, Nature \textbf{410}, 345 (2001).
% 14
\bibitem{Ji_APL06} Y. Ji, A. Hoffman, J.E. Pearson, and S.D. Bader, Appl. Phys. Lett. \textbf{88}, 052509 (2006).
% 15
\bibitem{Nogues_JMMM99} J. Nogu{\'e}s and I.K. Schuller, J. Magn. Magn. Mat. \textbf{192}, 203 (1999).
% 16
\bibitem{Rashba_APL02} E.I. Rashba, Appl. Phys. Lett. \textbf{80}, 2329 (2002).
% 17
\bibitem{Zhang_PRB02} S. Zhang and P.M. Levy, Phys. Rev. B \textbf{65}, 052409 (2002).
% 18
\bibitem{Dery_lateral_PRB06} H. Dery, {\L}. Cywi{\'n}ski, and L. J. Sham, Phys. Rev. B {\bf 73} 041306(R) (2006).
% 19
\bibitem{Yu_Flatte_long_PRB02} Z. G. Yu and M. E. Flatt{\'e}, Phys. Rev. B \textbf{66}, 235302 (2002).
% 20
\bibitem{Hershfield_PRB97} S. Hershfield and H. L. Zhao, Phys. Rev. B \textbf{56}, 3296 (1997).
% 21
\bibitem{Smith_Semiconductors} R. A. Smith, {\it Semiconductors} (Cambridge University Press, Cambridge, England, 1978).
% 22
\bibitem{Brataas} A. Brataas. Y. V. Nazarov and G. E. W. Bauer, Phys. Rev. Lett. {\bf 84}, 2481 (2000); Eur. Phys. J. B {\bf 22}, 99 (2001).
% 23
\bibitem{Rashba_PRB00}  E. I. Rashba, Phys. Rev. B \textbf{62}, R16267 (2000).
\end{thebibliography}
\end{document}